
\input amstex
\documentstyle{amsppt}
\overfullrule0pt
\def\dspl{\displaystyle}
\def\frc{\frac}
\def\qqqq#1#2{(#1,#2)\qquad}
\def\qqqt#1#2{(#1,#2)\hskip18pt}
\def\fmn#1#2{{F_{#1#2}}}
\def\nnnn{\vskip8pt}
\def\nnc{\vskip5pt\strut\hskip4pc}
\def\nnp{\vskip3pc}
\leftline{ hep-th/9506172 }
\vskip2pc
\topmatter
\title  The equations of some dispersionless limit
\endtitle
\author Seung Hwan Son
\endauthor
\abstract These equations are the generalized equations
of several dispersionless equations.
A complete table for $p\leq 10$ is provided.
\endabstract
\endtopmatter
\document
\head 1. Introduction
\endhead
\footnotetext""{1991 {\it Mathematics Subject Classification.}
Primary 35Q53, 35Q51.}
\footnotetext""{{\it Key words and phrases.}
Dispersionless limit, KdV, KP, Boussinesq equation.}
It is well-known that a lot of nonlinear solitonic equations
can be transformed into certain Hirota type bilinear equations
[17].
The $\tau$-function of the KP hierarchy can be characterized by the
Hirota equations and the Pl\"ucker relations are given from these
equations. The differential Fay identity which has quasi-classical
limit, is a part of the Pl\"ucker relations. The leading term of
the quasi-classical limit of the differential Fay identity satisfies
an identity [22] and from the identity
EQUATION($\cdot,\infty$)
is extracted.${}^*$
\footnotetext"* (Caution)"{
\item{} This notation is used only for technical
simplicity of expression since the formula was found by the
author recently (Fall, 1994) [5].
This is another way of using equation
numbers. Therefore, the author
strongly recommends careful use of the notation until
the formula is well-known. (Of course, given notation
has no meaning elsewhere like other usual equation numbers.)}
 Therefore, EQUATION($\cdot,\infty$)
is a subset of the dispersionless KP hierarchy.
EQUATION($p,q$) are derived from
EQUATION($p,\infty$). EQUATION($\cdot,2$) can be regarded
as a subset of dispersionless KdV hierarchy. We can
easily show that EQUATION($4,3$) is a dispersionless
Boussinesq equation. Therefore, EQUATION($\cdot,3$) can
be regarded as a subset of dispersionless  Boussinesq
hierarchy.
EQUATION($\cdot,q$) for $q>3$ is a whole new set of
dispersionless equations which can be regarded as a
subset of new hierarchy which may have some useful application.

\head 2. The formula
\endhead
Let us use $F_{mn}$ instead of $\displaystyle\frac{\partial^2}
{\partial t_m\partial t_n}\left(
F(t_1,\ldots,t_r,\ldots)\right)$.
\vskip10pt
\noindent{\bf Definition 2.1.} \qquad EQUATION($p,q$):
\hfill\break
${\dspl\sum_{0<i_1<\cdots<i_{k_p}\atop (i_1+1)n_{i_1}
+\cdots+(i_{k_p}+1)n_{i_{k_p}}=p}}
\left(\Bigl(\sum\limits_{j=1}^{k_p} n_{i_j}-1\Bigr)!
\prod\limits_{j=1}^{k_p}{\displaystyle
 \frc{(-\,F_{1i_j})^{n_{i_j}}}
{n_{i_j}!}}\right)$\hfill\break
\strut \hskip14pc
$+{\dspl\sum_{m+n=p} \frc{F_{mn}}{mn}}=0.$
\hfill\break
where the terms having $\frac{\partial}{\partial t_{q}},\ldots,
\frac{\partial}{\partial t_{kq}},\ldots$ vanish.

\head 3. EQUATION($4,q$)
\endhead
One can easily show that
EQUATION(4,$\infty$) is a dispersionless KP and
EQUATION(4,2) is a dispersionless KdV.

Consider a dispersionless Boussinesq equation
$$(uu_x)_x+\frac12u_{yy}=0.\eqno(3.1)$$
If we set $t_1=x$ and $t_2=y$, then EQUATION(4,3) is
$$\frac12(F_{xx})^2+\frac14F_{yy}=0.\eqno(3.2)$$
Differentiate (3.2) with respect to $x$.
Then we get
$$F_{xx}F_{xxx}+\frac14F_{xyy}=0.\eqno(3.3)$$
Setting $u=2F_{xx}$, (3.3) becomes
$$\left(\frac{u}2\frac{u_x}2\right)_x+
\frac14\left(\frac{u_{yy}}2\right)=0.$$
which is the same as (3.1).

\head 4. Discussion
\endhead

We could get the useful expression of the generalized equations
of the dispersionless limit of KdV, KP and Boussinesq equations.
And one can get a specific equation for each $(p,q)$.
Furthermore, new hierarchies are derived from EQUATION($\cdot,q$)
for $q>3$.
For further research, a table of equations are
provided.
By definition, EQUATION($p,q$) is the same as EQUATION($p,\infty$)
for $q\geq p$.
\vskip4pc

\head Table.\qquad Equations for $(p,q)$.
\endhead
\vskip1pc

\qqqq4{$\infty$}
$\frc12\fmn11^2-\frc13\fmn13+\frc14\fmn22=0$\nnnn
\qqqq5{$\infty$}
$\fmn11\fmn12-\frc12\fmn14+\frc13\fmn23=0$\nnnn
\qqqq6{$\infty$}
$\frc13\fmn11^3-\frc12\fmn12^2-\fmn11\fmn13+\frc35\fmn15-\frc19\fmn33-
\frc14\fmn24=0$\nnnn
\qqqq7{$\infty$}
$\fmn11^2\fmn12-\fmn12\fmn13-\fmn11\fmn14+\frc23\fmn16-\frc16\fmn34
-\frc15\fmn25=0$\nnnn
\qqqq8{$\infty$}
$\frc14\fmn11^4-\fmn11\fmn12^2-\fmn11^2\fmn13+\frc12\fmn13^2
+\fmn12\fmn14+\fmn11\fmn15$\nnc
$-\frc57\fmn17+\frc1{16}\fmn44+\frc2{15}\fmn35
+\frc16\fmn26=0$\nnnn
\qqqq9{$\infty$}
$\fmn11^3\fmn12-\frc13\fmn12^3-2\fmn11\fmn12\fmn13-\fmn11^2\fmn14
+\fmn13\fmn14$\nnc
$+\fmn12\fmn15+\fmn11\fmn16-\frc34\fmn18+\frc1{10}\fmn45
+\frc19\fmn36+\frc17\fmn27=0$\nnnn
\qqqt{10}{$\infty$}
$\frc15\fmn11^5-\frc32\fmn11^2\fmn12^2-\fmn11^3\fmn13
+\fmn12^2\fmn13+\fmn11\fmn13^2$\nnc
$+2\fmn11\fmn12\fmn14
-\frc12\fmn14^2+\fmn11^2\fmn15-\fmn13\fmn15
-\fmn12\fmn16-\fmn11\fmn17$\nnc
$+\frc79\fmn19-\frc1{25}\fmn55
-\frc1{12}\fmn46-\frc2{21}\fmn37-\frc18\fmn28=0$
\nnp
\qqqq42
$\frc12\fmn11^2-\frc13\fmn13=0$\nnnn
\qqqq52
$0=0$\nnnn
\qqqq62
$\frc13\fmn11^3-\fmn11\fmn13+\frc35\fmn15
-\frc19\fmn33=0$\nnnn
\qqqq72
$0=0$\nnnn
\qqqq82
$\frc14\fmn11^4-\fmn11^2\fmn13+\frc12\fmn13^2
+\fmn11\fmn15-\frc57\fmn17+\frc2{15}\fmn35=0$\nnnn
\qqqq92
$0=0$\nnnn
\qqqt{10}2
$\frc15\fmn11^5-\fmn11^3\fmn13
+\fmn11\fmn13^2+\fmn11^2\fmn15-\fmn13\fmn15
-\fmn11\fmn17$\nnc
$+\frc79\fmn19-\frc1{25}\fmn55
-\frc2{21}\fmn37=0$
\nnp
\qqqq43
$\frc12\fmn11^2+\frc14\fmn22=0$\nnnn
\qqqq53
$\fmn11\fmn12-\frc12\fmn14=0$\nnnn
\qqqq63
$\frc13\fmn11^3-\frc12\fmn12^2+\frc35\fmn15-
\frc14\fmn24=0$\nnnn
\qqqq73
$\fmn11^2\fmn12-\fmn11\fmn14
-\frc15\fmn25=0$\nnnn
\qqqq83
$\frc14\fmn11^4-\fmn11\fmn12^2
+\fmn12\fmn14+\fmn11\fmn15-\frc57\fmn17+\frc1{16}\fmn44
=0$\nnnn
\qqqq93
$\fmn11^3\fmn12-\frc13\fmn12^3-\fmn11^2\fmn14
+\fmn12\fmn15-
\frc34\fmn18+\frc1{10}\fmn45+\frc17\fmn27=0$\nnnn
\qqqt{10}3
$\frc15\fmn11^5-\frc32\fmn11^2\fmn12^2
+2\fmn11\fmn12\fmn14
-\frc12\fmn14^2+\fmn11^2\fmn15
-\fmn11\fmn17$\nnc
$-\frc1{25}\fmn55
-\frc18\fmn28=0$
\nnp
\qqqq54
$\fmn11\fmn12+\frc13\fmn23=0$\nnnn
\qqqq64
$\frc13\fmn11^3-\frc12\fmn12^2-\fmn11\fmn13+\frc35\fmn15-
\frc19\fmn33
=0$\nnnn
\qqqq74
$\fmn11^2\fmn12-\fmn12\fmn13+\frc23\fmn16
-\frc15\fmn25=0$\nnnn
\qqqq84
$\frc14\fmn11^4-\fmn11\fmn12^2-\fmn11^2\fmn13+\frc12\fmn13^2
+\fmn11\fmn15-\frc57\fmn17+\frc2{15}\fmn35
+\frc16\fmn26=0$\nnnn
\qqqq94
$\fmn11^3\fmn12-\frc13\fmn12^3-2\fmn11\fmn12\fmn13
+\fmn12\fmn15+\fmn11\fmn16-
+\frc19\fmn36+\frc17\fmn27=0$\nnnn
\qqqt{10}4
$\frc15\fmn11^5-\frc32\fmn11^2\fmn12^2-\fmn11^3\fmn13
+\fmn12^2\fmn13+\fmn11\fmn13^2
+\fmn11^2\fmn15$\nnc
$-\fmn13\fmn15
-\fmn12\fmn16-\fmn11\fmn17+\frc79\fmn19-\frc1{25}\fmn55
-\frc2{21}\fmn37=0$
\nnp
\qqqq65
$\frc13\fmn11^3-\frc12\fmn12^2-\fmn11\fmn13-\frc19\fmn33-
\frc14\fmn24=0$\nnnn
\qqqq75
$\fmn11^2\fmn12-\fmn12\fmn13-\fmn11\fmn14+\frc23\fmn16-\frc16\fmn34
=0$\nnnn
\qqqq85
$\frc14\fmn11^4-\fmn11\fmn12^2-\fmn11^2\fmn13+\frc12\fmn13^2
+\fmn12\fmn14-\frc57\fmn17+\frc1{16}\fmn44
+\frc16\fmn26=0$\nnnn
\qqqq95
$\fmn11^3\fmn12-\frc13\fmn12^3-2\fmn11\fmn12\fmn13-\fmn11^2\fmn14
+\fmn13\fmn14+\fmn11\fmn16$\nnc
$-\frc34\fmn18
+\frc19\fmn36+\frc17\fmn27=0$\nnnn
\qqqt{10}5
$\frc15\fmn11^5-\frc32\fmn11^2\fmn12^2-\fmn11^3\fmn13
+\fmn12^2\fmn13+\fmn11\fmn13^2+2\fmn11\fmn12\fmn14$\nnc
$-\frc12\fmn14^2
-\fmn12\fmn16-\fmn11\fmn17+\frc79\fmn19
-\frc1{12}\fmn46-\frc2{21}\fmn37-\frc18\fmn28=0$
\nnp
\qqqq76
$\fmn11^2\fmn12-\fmn12\fmn13-\fmn11\fmn14-\frc16\fmn34
-\frc15\fmn25=0$\nnnn
\qqqq86
$\frc14\fmn11^4-\fmn11\fmn12^2-\fmn11^2\fmn13+\frc12\fmn13^2
+\fmn12\fmn14+\fmn11\fmn15-\frc57\fmn17+\frc1{16}\fmn44$\nnc
$+\frc2{15}\fmn35
=0$\nnnn
\qqqq96
$\fmn11^3\fmn12-\frc13\fmn12^3-2\fmn11\fmn12\fmn13-\fmn11^2\fmn14
+\fmn13\fmn14+\fmn12\fmn15$\nnc
$-\frc34\fmn18+\frc1{10}\fmn45+\frc17\fmn27=0$\nnnn
\qqqt{10}6
$\frc15\fmn11^5-\frc32\fmn11^2\fmn12^2-\fmn11^3\fmn13
+\fmn12^2\fmn13+\fmn11\fmn13^2+2\fmn11\fmn12\fmn14$\nnc
$-\frc12\fmn14^2+\fmn11^2\fmn15-\fmn13\fmn15
-\fmn11\fmn17+\frc79\fmn19-\frc1{25}\fmn55
-\frc2{21}\fmn37-\frc18\fmn28=0$
\nnp
\qqqq87
$\frc14\fmn11^4-\fmn11\fmn12^2-\fmn11^2\fmn13+\frc12\fmn13^2
+\fmn12\fmn14+\fmn11\fmn15$\nnc
$+\frc1{16}\fmn44+\frc2{15}\fmn35
+\frc16\fmn26=0$\nnnn
\qqqq97
$\fmn11^3\fmn12-\frc13\fmn12^3-2\fmn11\fmn12\fmn13-\fmn11^2\fmn14
+\fmn13\fmn14+\fmn12\fmn15$\nnc
$+\fmn11\fmn16-\frc34\fmn18+\frc1{10}\fmn45+\frc19\fmn36=0$
\nnnn
\qqqt{10}7
$\frc15\fmn11^5-\frc32\fmn11^2\fmn12^2-\fmn11^3\fmn13
+\fmn12^2\fmn13+\fmn11\fmn13^2+2\fmn11\fmn12\fmn14$\nnc
$-\frc12\fmn14^2+\fmn11^2\fmn15-\fmn13\fmn15
-\fmn12\fmn16+\frc79\fmn19-\frc1{25}\fmn55
-\frc1{12}\fmn46-\frc18\fmn28=0$
\nnp
\qqqq98
$\fmn11^3\fmn12-\frc13\fmn12^3-2\fmn11\fmn12\fmn13-\fmn11^2\fmn14
+\fmn13\fmn14+\fmn12\fmn15$\nnc
$+\fmn11\fmn16+\frc1{10}\fmn45+\frc19\fmn36+\frc17\fmn27=0$\nnnn
\qqqt{10}8
$\frc15\fmn11^5-\frc32\fmn11^2\fmn12^2-\fmn11^3\fmn13
+\fmn12^2\fmn13+\fmn11\fmn13^2+2\fmn11\fmn12\fmn14$\nnc
$-\frc12\fmn14^2+\fmn11^2\fmn15-\fmn13\fmn15
-\fmn12\fmn16-\fmn11\fmn17+\frc79\fmn19-\frc1{25}\fmn55$\nnc
$-\frc1{12}\fmn46-\frc2{21}\fmn37=0$
\nnp
\qqqt{10}9
$\frc15\fmn11^5-\frc32\fmn11^2\fmn12^2-\fmn11^3\fmn13
+\fmn12^2\fmn13+\fmn11\fmn13^2+2\fmn11\fmn12\fmn14$\nnc
$-\frc12\fmn14^2+\fmn11^2\fmn15-\fmn13\fmn15
-\fmn12\fmn16-\fmn11\fmn17-\frc1{25}\fmn55
-\frc1{12}\fmn46$\nnc
$-\frc2{21}\fmn37-\frc18\fmn28=0$
\nnp

\head References
\endhead
\roster
\item"[1]"
M. J. Ablowitz and H. Segur,
{\it Solitons and the Inverse Scattering Transform}, SIAM (1981).
\item"[2]"
S. Aoyama and Y. Kodama,
{\it A generalized Sato equation and the $W_\infty$ algebra},
Phys. Lett. B {\bf 278} (1992), 56--62.
\item"[3]"
S. Aoyama and Y. Kodama,
{\it The $M$-truncated KP hierarchy and matrix models},
Phys. Lett. B {\bf 295} (1992), 190--198.
\item"[4]"
S. Aoyama and Y. Kodama,
{\it Topological Conformal Field Theory with a Rational
$W$ Potential and the dispersionless KP hierarchy},
Mod. Phy. Lett. A, {\bf 9} No. 27 (1994), 2481--2492.
\item"[5]"
R. W. Carroll,
{\it On dispersionless Hirota type equations},
hep-th/9410063.
\item"[6]"
R. Dijkgraaf, H. Verlinde and E. Verlinde,
{\it Loop Equations and Virasoro Constraints in Non-Perturbative
2d Quantum Gravity}, Nucl. Phys. {\bf B348} (1991), 435--456.
\item"[7]"
R. Dijkgraaf, H. Verlinde and E. Verlinde,
{\it Topological Strings in $d<1$}, Nucl. Phys. {\bf B352} (1991),
59--86.
\item"[8]"
J. Gibbons and Y. Kodama,
{\it Solving dispersionless Lax Equations},
Singular Limits of Dispersive Waves, (Plenum, 1994), 61--66.
\item"[9]"
Y. Kodama and J. Gibbons,
{\it A Method for Solving the dispersionless KP hierarchy
and its Exact Solutions. II},
Phys. Lett. A {\bf 135} No. 3 (1989), 167--170.
\item"[10]"
Y. Kodama and J. Gibbons,
{\it Integrability of the dispersionless KP hierarchy},
Proc. of the Workshop on Nonlinear Processes in Physics
(World Scientific, 1990), p. 166.
\item"[11]"
Y. Kodama,
{\it Exact Solutions of Hyperdynamic type Equations having
Infinitely Many Conserved Densities},
Phys. Lett. A {\bf 135} No. 3 (1989), 171--174.
\item"[12]"
B. G. Konopelchenko,
{\it Introduction to Multidimensional Integrable Equations},
Plenum Press, (1992).
\item"[13]"
I. M. Krichever,
{\it Method of Averaging for Two-dimensional Integrable Equations},
Funct. Anal. Appl. {\bf 22} (1988), 200--213.
\item"[14]"
I. M. Krichever,
{\it The dispersionless Lax Equations and Topological Minimal Models},
Commun. Math. Phys. {\bf 143} (1991), 415--426.
\item"[15]"
I. M. Krichever,
{\it The $\tau$-Function of the Universal Whitham Hierarchy,
Matrix Models and Topological Field Theorie}, hep-th/9205110.
\item"[16]"
Y. Ohta, J. Satsuma, D. Takahashi and T. Tokihiro,
{\it An Elementary Introduction to Sato Theory}, Prog. Theoret.
Phys. Supp., {\bf 94} (1988), 210.
\item"[17]"
S. Oishi, {\it A Method of Analysing Soliton Equations by
Bilinearization,} J. Phys. Soc. Jpn, {\bf 48} (1980), 639.
\item"[18]"
M. Sato and Y. Sato,
{\it Soliton Equations as Dynamical Systems on Infinite Dimensional
Grassmann Manifold}, Proc. the U.S.-Japan Seminar, Tokyo, 1982.
\item"[19]"
M. Sato and M. Noumi,
{\it Soliton Equations and the Universal Grassmann Manifolds},
Sophia Univ. Kokyuroku in Math., {\bf 18} (1984).
\item"[20]"
K. Takasaki and T. Takebe,
{\it SDiff(2) KP Hierarchy},
Int. J. Mod. Phys. {\bf A7, Suppl. 1B} (1992), 889--922.
\item"[21]"
K. Takasaki and T. Takebe,
{\it Quasi-Classical Limit of KP Hierarchy, W-Symmetries and
Free Fermions}, Kyoto preprint KUCP-0050/92 (July, 1992).
\item"[22]"
K. Takasaki and T. Takebe,
{\it Integrable hierarchies and dispersionless limit},
hep-th/940596.
\item"[23]"
E. Witten,
{\it Ground Ring of Two Dimensional String Theory},
Nucl. Phys. {\bf B373} (1992), 187--213.
\endroster
\vskip2pc
{\obeylines
Seung H. Son
Department of Mathematics
University of Illinois at Urbana-Champaign
{\it E-mail address}:  {\tt son\@math.uiuc.edu}
}
\enddocument
\bye